\documentclass[iop]{emulateapj}

\usepackage{ulem}
\usepackage{bm}
\usepackage{color}
\usepackage{verbatim}
\usepackage{lineno}

\newcommand\kms{\ifmmode{\rm km\thinspace s^{-1}}\else km\thinspace s$^{-1}$\fi}
\newcommand\jaavso{Journal of the American Association of Variable Star Observers}

\shortauthors{Torres}
\shorttitle{Polaris}

\begin{document} 
\submitted{Accepted for publication in Monthly Notices of the Royal Astronomical Society}
%\linenumbers
\title{The Spectroscopic Orbit of Polaris, and its pulsation Properties}

\author{
Guillermo Torres
}

\affil{Center for Astrophysics $\vert$ Harvard \&
  Smithsonian, 60 Garden St., Cambridge, MA 02138, USA;
  gtorres@cfa.harvard.edu}
  
\begin{abstract}
Polaris is the nearest and brightest classical Cepheid, and pulsates
with a period of about 4 days. It has long been known as a
single-lined spectroscopic binary with an orbital period of
30~yr. Historical photometric and spectroscopic records indicate that,
until recently, the pulsation period has been increasing at a rate of
about 4.5~s~yr$^{-1}$, and that the amplitude of the pulsation
declined for most of the 20th century, but more recently halted its
decline and began to increase. Here we report an analysis of the more
than 3600 individual radial velocity measurements of Polaris available
from the literature over the past 126~yr. We find that the pulsation
period is now becoming shorter, and that the amplitude of the velocity
variations has stopped increasing, and may be getting smaller
again. We also find tantalising evidence that these changes in
pulsation behaviour over the last century may be related to the binary
nature of the system, as they seem to occur near each periastron
passage, when the secondary comes within 29 stellar radii of the
Cepheid in its eccentric orbit. This suggests the companion may be
perturbing the atmosphere of the Cepheid and altering its pulsation
properties at each encounter.  After removal of the pulsation
component of the velocities, we derive a much improved spectroscopic
orbit for the binary that should serve as the basis for a more
accurate determination of the dynamical masses, which are still rather
uncertain.
\end{abstract}


%%%%%%%%%%%%%%%%%%%%%%%%%%%%%%%%%%%%%%%%%%%%%%%%%%%%%%%%%%%%%%%%%%%%%
\section{Introduction}
\label{sec:introduction}
%%%%%%%%%%%%%%%%%%%%%%%%%%%%%%%%%%%%%%%%%%%%%%%%%%%%%%%%%%%%%%%%%%%%%

Polaris ($\alpha$~UMi), the North Star, needs little introduction.
Because of its brightness and location only 45\arcmin\ from the north
celestial pole, it has been an invaluable navigation beacon for
centuries. It has also been the subject of numerous scientific
investigations since the mid 1850's, when it's brightness was
discovered to vary in a periodic fashion.  Polaris is a yellow
supergiant (\ion{F7}{1}b) and a classical Cepheid, albeit a somewhat
unusual one for its short period (very close to 4~days), low amplitude
(currently about 0.07~mag in $V$), and other characteristics of its
variability. It is considered to be a first overtone pulsator
\citep[see, e.g.,][]{Feast:1997, Evans:2002, Neilson:2012}.

Polaris is also a triple system. It is attended by an
18\arcsec\ companion (Polaris~B, \ion{F3}{5}) discovered in 1779 by
William Herschel \citep{Herschel:1782}, and by a much closer
spectroscopic companion (Polaris~Ab, \ion{F6}{5}) found more than a
century later by observers at the Lick Observatory, and first
mentioned by \cite{Campbell:1899}.\footnote{At the time of Campbell's
  report, the more obvious 4-day radial velocity variability seen by
  the Lick astronomers, and others, was thought to be caused by an
  additional spectroscopic companion. The true pulsation nature of
  that 4-day oscillation was realised some years later.}  This latter
companion, with its orbital period of about 30~yr, is one of the
subjects of this paper. In the context of its multiplicity, we will
refer to the Cepheid itself as Polaris~Aa.

As the closest Cepheid to the Earth, as well as the brightest ($V =
2.00$), Polaris is an important laboratory for understanding the
Cepheid phenomenon, and yet its most fundamental property ---its
mass--- remains poorly known. This is all the more unfortunate given
the longstanding discrepancy between masses predicted by stellar
evolution models and by pulsation calculations, and the still few
precise empirical mass measurements available for Cepheids \citep[see,
  e.g.,][]{Bono:2001, Neilson:2011}.  The most recent dynamical mass
estimate for Polaris~Aa, relying on the binary nature of the object,
is $3.45 \pm 0.75~M_{\sun}$ \citep{Evans:2018}.  This determination
was made possible by UV imaging observations with the Hubble Space
Telescope (HST) that resolved the 30~yr spectroscopic companion for
the first time \citep{Evans:2008, Evans:2018}, and by an improved
parallax for the system from the Gaia DR2 catalogue \citep{Gaia:2018}.

While the radial velocity (RV) of Polaris has been monitored for a
very long time, the derivation of the spectroscopic orbit has always
been complicated by the fact that changes in the velocity due to the
pulsation are superimposed on the long-period orbital motion, and are
of comparable magnitude. It would be a relatively simple matter to
account for this if the oscillations were regular, but in Polaris they
are not. Both the amplitude and the period of the pulsation are
changing, slowly but in an irregular and unpredictable manner over
timescales of decades, or perhaps shorter. As of the most recent study
from about a decade ago \citep{Neilson:2012}, the pulsation period had
been increasing monotonically for more than a century and a half at a
rapid rate of about 4 or 5 seconds per year, though with an apparent
``glitch" in the mid 1960s \citep{Turner:2005}.  More dramatic changes
have been seen in the amplitude of the pulsation.  Photometric and
radial velocity measurements showed little or no change in the
amplitude for most of the 20th century, but then a sharp decline was
observed after about 1980 \citep{ArellanoFerro:1983a, Kamper:1984,
  Dinshaw:1989, Brown:1994}. Combined with the lengthening of the
period, this drop in the amplitude fuelled speculation in the early
1990s that it might presage the complete cessation of the pulsations
by 1994 or 1995 \citep{Dinshaw:1989, Fernie:1993}, signaling the
evolution of the star out of the instability strip. This did not come
to pass, however, as the pulsation amplitude first held steady at a
low level in the 1990s \citep{Kamper:1998, Hatzes:2000}, and then
began to increase again after the year 2000 \citep{Bruntt:2008,
  Spreckley:2008, Lee:2008}. This unusual behaviour has attracted much
attention over the last decade or so, and prompted detailed attempts
to model the evolution of the Cepheid, but remains largely
unexplained.  The classical Cepheid HDE~344787, considered by some to
be an analog of Polaris \citep{Turner:2010, Ripepi:2021}, shares many
of its pulsation properties including a decreasing amplitude ---now
barely discernible except from space--- and a rapidly increasing
period.  The rate of period change, however, is nearly 3 times faster
than Polaris.  It has also been predicted to stop pulsating in the not
too distant future.

In one of the more influential spectroscopic studies of Polaris,
\cite{Kamper:1996} combined an extensive series of RV measurements
from the David Dunlap Observatory (1980--1995) with earlier ones from
the Lick Observatory \citep[1896--1958;][]{Roemer:1965}, and derived
elements for the 30~yr single-lined spectroscopic orbit that have been
widely used by many researchers since. The pulsation component of the
velocities was removed from the original observations by subtracting
sine curve fits from suitably grouped subsets of data.  Subsequent
studies by others have applied similar techniques, but have only made
use of some of the available data sources to derive revised
spectroscopic elements.

An exhaustive search of the literature since Kamper's 1996 paper, and
also prior to that time, has identified several new RV samples, both
published and unpublished, which have not been used for the purpose of
orbit determination, or to reexamine the pulsation properties. They
extend the baseline up to the present, more than doubling the number
of individual measurements. This new information makes it timely to
return to Polaris to investigate its present status.

The structure of the paper is as follows. In Section~\ref{sec:rvs} we
briefly discuss the available sources of RVs for Polaris. Details of
the various datasets are collected in the
Appendix. Section~\ref{sec:analysis} gives a description of the
treatment of the velocities, and how we disentangled the pulsations
from the orbital motion. The results of our analysis are reported in
Section~\ref{sec:results}, beginning with the derivation of much
improved elements of the spectroscopic orbit
(subsection~\ref{sec:orbit}). Next we present our study of the
$O\!-\!C$ diagram, based on the RV measurements as well as on existing
and new times of maximum light, revealing how the pulsation period has
changed over the last 175~yr (subsection~\ref{sec:o-c}). This is
followed in Section~\ref{sec:amplitude} with our results on the
changes that have occurred in the pulsation
amplitude. Section~\ref{sec:discussion} discusses the above results in
the context of the binarity of Polaris. We conclude with final
thoughts in Section~\ref{sec:conclusions}.

%%%%%%%%%%%%%%%%%%%%%%%%%%%%%%%%%%%%%%%%%%%%%%%%%%%%%%%%%%%%%%%%%%%%%
\section{Radial Velocity Material}
\label{sec:rvs}
%%%%%%%%%%%%%%%%%%%%%%%%%%%%%%%%%%%%%%%%%%%%%%%%%%%%%%%%%%%%%%%%%%%%%

The earliest RV measurements of Polaris of which we are aware date
back to 1888 \citep{Vogel:1895}. Since then, more than two dozen
publications have reported spectroscopic observations, the most
extensive of which is the study by \cite{Roemer:1965}, featuring
nearly 1200 measurements made at the Lick Observatory over more than
60~yr.  Some of the other datasets are too scattered or too poor in
quality to be useful.  Others have been superseded, partially or
entirely, by subsequent publications in which the observations were
re-reduced or their uncertainties or zeropoints changed.
Table~\ref{tab:historical.rvs} lists all of the RV sources we have
identified, with their Julian date range, number of original
measurements, and the wavelength range of the observations, when
available.  They are sorted by the date of the first observation. In
two cases the original measurements are unpublished, and we have
obtained them from the authors. In three others, they are unavailable.

\setlength{\tabcolsep}{6pt}
\begin{deluxetable*}{lcccl}
\tablecaption{Sources of Radial Velocities for Polaris \label{tab:historical.rvs}}
\tablehead{
\colhead{Source} &
\colhead{Date Range} &
\colhead{Time Span} &
\colhead{$N_{\rm obs}$} &
\colhead{Wavelength Range}
\\
\colhead{} &
\colhead{(JD$-$2,400,000)} &
\colhead{(days)} &
\colhead{} &
\colhead{}
}
\startdata
%                                                  JD              Span         N        Wavelength range
\cite{Vogel:1895}                           &  10956--10978      &     22  &      2    &  Photographic, H$\gamma$ region \\
\cite{Roemer:1965}                          &  13811--36505      &  22694  &   1180    &  Photographic, H$\gamma$ region before 1903, 4500~\AA\ afterwards \\
\cite{Frost:1899}                           &  14876--14924      &     48  &      3    &  Photographic \\
\cite{Belopolsky:1900}                      &  14982--15109      &    127  &     17    &  Photographic, H$\gamma$ region \\
\cite{Hartmann:1901}                        &  15086--15403      &    317  &     35    &  Photographic \\
\cite{Kustner:1908}                         &  16298--17016      &    718  &      7    &  Photographic, 4150--4500~\AA \\
\cite{Abt:1970}                             &  20997--27024      &   6027  &      3    &  Photographic, H$\gamma$ region \\
\cite{Henroteau:1924}                       &  23539--23676      &    137  &     58    &  Photographic, 4000--4600~\AA \\
\cite{Schmidt:1974}                         &     40971          &      0  &      1    &  Photographic (Kodak 098-02), H$\alpha$ region \\
\cite{Wilson:1989}                          &  43297--43300      &      3  &      3    &  3850--4250~\AA \\
\cite{ArellanoFerro:1983a}\tablenotemark{a} &  44432--44988      &    556  &     35    &  Photographic, mostly 4000--5000~\AA, some 5300--6700~\AA \\
\cite{Beavers:1986}                         &     44449          &      0  &      1    &  650~\AA\ region centred at 4600~\AA \\
\cite{Kamper:1984}\tablenotemark{b}         &  45414--45592      &    178  &     33    &  Photographic (IIaO, IIIaJ) \\
\cite{Kamper:1996}\tablenotemark{c}         &  45414--49600      &   4186  &    307    &  Photographic (IIaO, IIIaJ), Reticon/CCD \\
\cite{Dinshaw:1989}                         &  46922--47163      &    241  &    175    &  4220--4700~\AA \\
\cite{Sasselov:1990}                        &  47432--47436      &      4  &      2    &  350~\AA\ region centred at 1.1~$\mu$m \\
\cite{Gorynya:1992}                         &  47962--48415      &    453  &     32    &  4000--6000~\AA \\
\cite{Garnavich:1993}\tablenotemark{d}      &   1992--           & \nodata &  \nodata  &  Hydrogen fluoride cell, 8600~\AA \\
\cite{Hatzes:2000}                          &  48581--49201      &    620  &     42    &  23.6~\AA\ region centred at 5520~\AA \\
\cite{Gorynya:1998}                         &  49446--49569      &    123  &     40    &  4000--6000~\AA \\
\cite{Usenko:2015}                          &  49512--54934      &   5422  &     56    &  5800--6800~\AA, 5500--7000~\AA, 4470--7100~\AA \\
\cite{Kamper:1998}                          &  49950--50568      &    618  &    212    &  6290~\AA\ after early 1994 \\
\cite{Eaton:2020}                           &  52998--55138      &   2140  &    679    &  5000--7100~\AA \\
\cite{Lee:2008}\tablenotemark{e}            &  53332--54269      &    937  &    265    &  $I_2$ cell, $\sim$5000--6000~\AA \\
\cite{Bucke:2021}\tablenotemark{e}          &  53482--60005      &   6523  &    296    &  6050--6700~\AA \\
\cite{Fagas:2009}\tablenotemark{f}          &  Dec2007--Jul2008  & \nodata &    330    &  \nodata \\
\cite{deMedeiros:2014}\tablenotemark{f}     &   \nodata          &   6098  &     13    &  3600--5200~\AA \\
\cite{Anderson:2019}                        &  55816--58443      &   2627  &    161    &  4813--6476~\AA \\
\cite{Usenko:2016}                          &  57283--57376      &     93  &     21    &  4900--6800~\AA \\
\cite{Usenko:2017}                          &  57623--57816      &    193  &     49    &  4900--6800~\AA \\
\cite{Usenko:2018}                          &  57971--58248      &    277  &     63    &  4900--6800~\AA \\
\cite{Usenko:2020}                          &  58368--58955      &    587  &     53    &  4900--6800~\AA 
\enddata

\tablecomments{Measurements by \cite{Vogel:1895}, \cite{Frost:1899}, \cite{Abt:1970},
\cite{Schmidt:1974}, \cite{Wilson:1989}, \cite{Beavers:1986}, and \cite{Sasselov:1990} are either
poor, too few, or too scattered to be useful.}
\tablenotetext{a}{Most of these measurements were superseded by the publication of \cite{Kamper:1996}, in which
a +0.5~\kms\ offset was applied.}
\tablenotetext{b}{All of these measurements were superseded by the publication of \cite{Kamper:1996}.}
\tablenotetext{c}{Some of these measurements were superseded by the publication of \cite{Kamper:1998}.}
\tablenotetext{d}{No other details or individual RVs were published.}
\tablenotetext{e}{Unpublished measurements kindly provided by the author.}
\tablenotetext{f}{Unpublished. Only the time interval or time span was reported.}
\end{deluxetable*}
\setlength{\tabcolsep}{6pt}

Several of the sources in Table~\ref{tab:historical.rvs} contain only
relative velocities. These can still be useful for constraining the
spectroscopic orbit of Polaris, provided they span a phase interval
during which the orbital velocity changed significantly relative to
the precision of the observations. A few of the datasets do not meet
this requirement, as discussed later, and have not been considered for
our analysis.  The velocities reported by \cite{Dinshaw:1989} are not
only measured relative to the first observation, but they additionally
have a quadratic approximation to the orbital motion removed. This
change can be easily undone, as we describe in the Appendix, where
further details of this and all other datasets are discussed. The same
holds for the velocities of \cite{Kamper:1998}, which are relative and
have the orbital motion subtracted. In addition to the dominant 4\,d
pulsation signal, a few of the studies have reported secondary
periodicities in the velocities \citep{Dinshaw:1989, Hatzes:2000,
  Lee:2008} or are affected by seasonal variations of instrumental
nature that need to be corrected \citep{Eaton:2020, Bucke:2021}. This
will be described in the next section.

\begin{figure*}
\epsscale{1.17}
\plotone{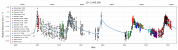}
\figcaption{Historical radial velocities of Polaris, shown with the
  orbital model by \cite{Kamper:1996} for reference. The large
  scatter is caused by the pulsation motion, which is superimposed on
  the orbital motion. Datasets
  reporting relative velocities have been shifted by eye to bring them
  in line with the others. \label{fig:raw.rvs}}
\end{figure*}

Figure~\ref{fig:raw.rvs} presents a visual summary of all of the
observational material, which now covers more than 4 orbital cycles
and contains 3727 individual RV measurements.  The large scatter is
caused mostly by the pulsations. As seen in the plot, there is a
complete lack of useful RV measurements for two full decades, from
1960 to 1980.\footnote{We have located only a handful of scattered RV
  measurements for Polaris from 1971 \citep{Schmidt:1974} and 1977
  \citep{Wilson:1989}, but they are too few to be of use here (see the
  Appendix, and Table~\ref{tab:historical.rvs}).}  This is rather
unfortunate, because it is precisely during this time when there were
significant changes in the characteristics of the pulsation, as we
will see later.

%%%%%%%%%%%%%%%%%%%%%%%%%%%%%%%%%%%%%%%%%%%%%%%%%%%%%%%%%%%%%%%%%%%%%
\section{Radial Velocity Analysis}
\label{sec:analysis}
%%%%%%%%%%%%%%%%%%%%%%%%%%%%%%%%%%%%%%%%%%%%%%%%%%%%%%%%%%%%%%%%%%%%%

Because the velocity amplitude and period of the pulsation are not
regular, removal of the 4\,d oscillations from the original RV data in
order to study the orbital motion requires us to determine the
properties of the pulsation as part of the process. The procedure we
followed is iterative, alternating between refining the pulsation
properties and improving the orbital solution. As is customary in the
study of variable stars, we relied on the $O\!-\!C$ diagram to
investigate changes in the pulsation period. This diagram typically
shows the difference between observed times of maximum light and
calculated times based on a fixed linear ephemeris. In this case, we
are dealing with radial velocities rather than brightness
measurements, but it has been found that the velocity minima for
small-amplitude pulsating yellow supergiants such as Polaris
correspond quite closely to light maxima, with only a modest time
shift \citep{ArellanoFerro:1983b}. The study of \cite{Turner:2005}
established the offset to be $-0.383$~d, which they added to their
measured times of minimum RV in order to align them with the times of
maximum light. A smaller adjustment of $-0.21$~d was advocated by
\cite{ArellanoFerro:1983a}, and \cite{Kamper:1998} adopted an
intermediate value of $-0.333$~d. As we show later in
Section~\ref{sec:o-c}, the value of $-0.21$~d seems more consistent
with our measurements, and is adopted for this paper.

For the present analysis we followed \cite{Turner:2005}, and others, in adopting
the ephemeris of \cite{Berdnikov:1995} for the times of maximum light:
\begin{equation}
\label{eq:ephemeris}
{\rm HJD}_{\rm max} = 2,\!428,\!260.727 + 3.969251 E~,
\end{equation}
where $E$ represents the number of cycles elapsed since the reference
epoch.

We began by subtracting from the original velocities the orbital
motion predicted from the elements reported by \cite{Kamper:1996}.
Datasets with relative velocities were offset by eye so as come close
to others near in time. All dates of observation were corrected for
light travel time in the 30~yr orbit to refer them to the barycentre
of the binary, based on the same set of orbital elements. In this case
those corrections are small, and range from $-0.025$~d to $+0.007$~d
throughout the orbit. As in most Cepheids of low amplitude, the
brightness and velocity variations in Polaris are very nearly
sinusoidal \citep[see, e.g.,][]{Efremov:1975}, which makes the process
of subtracting out the pulsation simpler. However, prior to that step,
we removed additional periodicities or other spurious variations in
some of the datasets so as to avoid extra scatter. For example,
\cite{Dinshaw:1989} reported a secondary period of about 45~d in their
data. In this case we chose to fit the sum of two sinusoids (for the
4\,d pulsation component, and the 45\,d component), allowing the
periods and amplitudes to vary, and then subtracted only the 45\,d
component from the orbit-corrected velocities, leaving just the
pulsations. We proceeded similarly with a 40\,d signal found by
\cite{Hatzes:2000}, and a 119\,d signal reported by
\cite{Lee:2008}. The \cite{Eaton:2020} RVs showed a fairly obvious
annual pattern in the residuals of unknown origin. We removed it in
the same way.\footnote{\cite{Anderson:2019} carried out a search for
  additional periodicities in their spectroscopic observations by
  examining the line bisectors, rather than the velocities, because
  this avoids complications from the orbital motion as the bisectors
  should be unaffected. Aside from the pulsations, they found evidence
  for a 40\,d variation (the same as in \citealt{Hatzes:2000}) and a
  60\,d variation, although they had reservations about the latter
  signal. While it is not a goal of this paper to perform a detailed
  search for other periodicities in the published RVs, our
  two-component sine fit to the \cite{Anderson:2019} velocities did
  not reveal a signal at 40\,d with an amplitude significant enough to
  affect our results, so for our purposes we have not considered that
  signal further.} The velocities of \cite{Bucke:2021} are affected by
seasonal variations related to thermal effects in their
instrumentation (see the Appendix). In this case, we first removed the
pulsation component and then fitted the residuals with a spline
function, and subtracted this spline fit from the orbit-corrected
data. This left only the changes due to pulsation.

With the orbital motion removed, and any additional periodicities also
removed as just described, we then divided each dataset into subsets
in time, and fitted a sine curve to the RVs in each interval to model
the pulsation. We solved for the amplitude of the pulsation variation
($A_{\rm puls}$), the time of velocity minimum nearest to the average
date of each interval ($T_{\rm min}$), and a velocity offset. The
period was initially held fixed at the value in
eq.[\ref{eq:ephemeris}]. The definition of these intervals,
particularly for some of the sources with sparse sampling, was a
compromise between having enough observations to sample the pulsation
cycle, and a time span short enough to not smear out real evolution in
the amplitude or phasing. The median number of observations per
interval is about two dozen, and the median duration is about two
months, although there is a significant range in both, depending on
the dataset.  Because the period is changing, this step of modelling
the pulsation with a sine function was repeated after the initial fit,
using the resulting $O\!-\!C$ diagram to predict a more accurate
period at the mean epoch of each interval, and holding it fixed.

The next step in our analysis was to subtract these sine curve fits
(and the additional periodicities mentioned earlier) from the original
RV data, which left only the orbital motion plus any shifts due to
differences in the velocity zeropoints of the various datasets. This
was then used to solve for the spectroscopic orbital elements, the
details of which we describe in the next section. The entire process
in this section was repeated several times until convergence, during
which a small number of RV outliers were rejected. We retained a total
of 3659 measurements.  As a result of this effort, we obtained
improved elements for the 30~yr orbit of Polaris, as well as a set of
homogeneous measurements that trace the evolution of the period and
velocity amplitude of the pulsations over more than a century. The
$O\!-\!C$ and RV amplitude measurements we obtained are presented in
Table~\ref{tab:omc}, along with other properties of the fits, and will
be discussed below.

\setlength{\tabcolsep}{4pt}
\begin{deluxetable*}{cccccccccc}
\tablecaption{Spectroscopic Amplitude and Timing Measurements for Polaris \label{tab:omc}}

\tablehead{
  \colhead{Mean JD} &
  \colhead{JD Interval} &
  \colhead{$N_{\rm obs}$} &
  \colhead{$E$} &
  \colhead{$A_{\rm puls}$} &
  \colhead{$T_{\rm min}$} &
  \colhead{$\Delta t$} &
  \colhead{$O\!-\!C$} &
  \colhead{RMS} &
  \colhead{Source}
  \\
  \colhead{(HJD$-$2,400,000)} &
  \colhead{(HJD$-$2,400,000)} &
  \colhead{} &
  \colhead{} &
  \colhead{(\kms)} &
  \colhead{(HJD$-$2,400,000)} &
  \colhead{(day)} &
  \colhead{(day)} &
  \colhead{(\kms)} &
  \colhead{}
  }
\startdata
% HJDmean       Interval        N         E               K                   Tminbary              tcor        O-C       RMS    Source  
13850.280  &  13812--13903  &   7  &  $-$3632  &  $2.506 \pm 0.170$  &  $13847.486 \pm 0.028$  &  $+0.005$  &  3.079  &  0.218  &  1 \\
14895.563  &  14876--14911  &  21  &  $-$3368  &  $2.882 \pm 0.153$  &  $14895.025 \pm 0.035$  &  $+0.002$  &  2.736  &  0.496  &  1 \\
14996.439  &  14945--15094  &  13  &  $-$3343  &  $3.283 \pm 0.228$  &  $14994.196 \pm 0.038$  &  $+0.001$  &  2.676  &  0.511  &  1 \\
15042.930  &  14988--15109  &  13  &  $-$3331  &  $1.944 \pm 0.350$  &  $15042.205 \pm 0.126$  &  \phs$0.000$  &  3.053  &  0.866  &  2 \\
15105.995  &  15086--15134  &  12  &  $-$3315  &  $2.497 \pm 0.838$  &  $15105.270 \pm 0.176$  &  $-0.001$  &  2.610  &  1.782  &  3 
\enddata

\tablecomments{Cycle numbers ($E$) are based on the ephemeris in
  eq.[\ref{eq:ephemeris}]. Column $A_{\rm puls}$ gives the velocity
  semiamplitude. Column $T_{\rm min}$ contains the measured times of
  RV minimum, corrected for the light travel time $\Delta t$ in the
  following column, and shifted by $-0.21$~d to line up with the times
  of maximum light (see the text). The $O\!-\!C$ residuals are computed from $T_{\rm
    min}$, the cycle number, and the ephemeris. The scatter of each
  sine curve fit is given in column RMS. Sources in the last column
  are as follows:
  (1) \cite{Roemer:1965};
  (2) \cite{Belopolsky:1900};
  (3) \cite{Hartmann:1901};
  (4) \cite{Kustner:1908};
  (5) \cite{Henroteau:1924};
  (6) \cite{ArellanoFerro:1983a};
  (7) \cite{Kamper:1996};
  (8) \cite{Dinshaw:1989};
  (9) \cite{Gorynya:1992};
  (10) \cite{Hatzes:2000};
  (11) \cite{Gorynya:1998};
  (12) \cite{Usenko:2015, Usenko:2016, Usenko:2017, Usenko:2018, Usenko:2020};
  (13) \cite{Kamper:1998};
  (14) \cite{Eaton:2020};
  (15) \cite{Lee:2008};
  (16) \cite{Bucke:2021};
  (17) \cite{Anderson:2019}.
  (This table is available in its entirety in machine-readable form)}

\end{deluxetable*}
\setlength{\tabcolsep}{6pt}

%%%%%%%%%%%%%%%%%%%%%%%%%%%%%%%%%%%%%%%%%%%%%%%%%%%%%%%%%%%%%%%%%%%%%
\section{Results}
\label{sec:results}
%%%%%%%%%%%%%%%%%%%%%%%%%%%%%%%%%%%%%%%%%%%%%%%%%%%%%%%%%%%%%%%%%%%%%

%%%%%%%%%%%%%%%%%%%%%%%%%%%%%%%%%%%%%%%%%%%%%%%%%%%%%%%%%%%%%%%%%%%%%
\subsection{Revised Spectroscopic Orbital Solution}
\label{sec:orbit}
%%%%%%%%%%%%%%%%%%%%%%%%%%%%%%%%%%%%%%%%%%%%%%%%%%%%%%%%%%%%%%%%%%%%%

The radial velocities of Polaris, modified as described above to
remove all contributions except for the orbital motion, span more than
126~yr, or 4.3 orbital cycles.  They are listed in
Table~\ref{tab:rvs}, together with their formal uncertainties.  The
measurements of \cite{ArellanoFerro:1983a} and \cite{Kamper:1996},
obtained at the same observatory, were merged and treated as a single
dataset for our analysis, after applying an offset of +0.5~\kms\ to
the former velocities to place them on the same reference frame as the
latter (see the Appendix). We also grouped together the RVs of
\cite{Gorynya:1992, Gorynya:1998}, which were made with the same
instrumentation and measuring technique.  Similarly, the RVs of Usenko
and collaborators, published in several batches over a period of 5~yr,
were combined and considered as one group \citep{Usenko:2015,
  Usenko:2016, Usenko:2017, Usenko:2018, Usenko:2020}. In total we
used 13 separate RV datasets.  The uncertainties were initially
adopted as published, and in cases where errors were not reported, we
assigned reasonable values. We refer the reader to the Appendix for
the details of each set.

\setlength{\tabcolsep}{3pt}
\begin{deluxetable}{cccccc}
\tablecaption{Radial Velocities for Polaris with the Pulsation Removed \label{tab:rvs}}
\tablehead{
\colhead{Date\tablenotemark{a}} &
\colhead{Year} &
\colhead{$\Delta t$} &
\colhead{RV} &
\colhead{$\sigma_{\rm RV}$} &
\colhead{Source}
\\
\colhead{(HJD)} &
\colhead{} &
\colhead{(day)} &
\colhead{(\kms)} &
\colhead{(\kms)} &
\colhead{}
}
\startdata
2413811.956  &  1896.6926  &  +0.004  &  $-17.706$  &  0.500  &  1 \\
2413818.953  &  1896.7117  &  +0.004  &  $-17.999$  &  0.500  &  1 \\
2413826.897  &  1896.7335  &  +0.005  &  $-17.907$  &  0.500  &  1 \\
2413838.878  &  1896.7663  &  +0.005  &  $-17.897$  &  0.500  &  1 \\
2413875.810  &  1896.8674  &  +0.005  &  $-18.054$  &  0.500  &  1 
\enddata

\tablecomments{The RVs in the table are listed as used in our
orbital analysis. The sources, which are numbered differently
than in Table~\ref{tab:omc} because some of the datasets were
merged for the orbital analysis and others were omitted, are as follows:
(1) \cite{Roemer:1965};
(2) \cite{Hartmann:1901};
(3) \cite{Kustner:1908};
(4) \cite{ArellanoFerro:1983a} + 0.5~\kms, and \cite{Kamper:1996};
(5) \cite{Dinshaw:1989};
(6) \cite{Gorynya:1992, Gorynya:1998};
(7) \cite{Hatzes:2000};
(8) \cite{Usenko:2015}, \cite{Usenko:2016}, \cite{Usenko:2017}, \cite{Usenko:2018}, and \cite{Usenko:2020};
(9) \cite{Kamper:1998};
(10) \cite{Eaton:2020};
(11) \cite{Lee:2008};
(12) \cite{Bucke:2021};
(13) \cite{Anderson:2019}.
The formal uncertainties listed are as published, or assigned here
when not published (see the Appendix). Those of \cite{Dinshaw:1989}
and Usenko were found to be overestimated, and have been reduced by a
factor of 5 from their reported values. For the final uncertainties
used in our orbital analysis,
see Table~\ref{tab:offsets}. (This table is available in its entirety
in machine-readable form)}

\tablenotetext{a}{Dates have been corrected for light travel time
to the barycentre of the Polaris~Aa+Ab system, using the orbital elements in
  Table~\ref{tab:mcmc}. The corrections $\Delta t$ are listed in the
  third column.}

\end{deluxetable}
\setlength{\tabcolsep}{6pt}

The six standard elements of a single-lined spectroscopic orbit were
solved with a Markov chain Monte Carlo (MCMC) procedure, using the
{\sc
  emcee}\footnote{\url{https://emcee.readthedocs.io/en/stable/index.html}}
package of \cite{Foreman-Mackey:2013}. The eccentricity and argument
of periastron were cast as $\sqrt{e}\cos\omega$ and
$\sqrt{e}\sin\omega$, in which $\omega$ is the value for the primary
(star Aa). Additionally, to account for differences in the velocity
zeropoints of the various datasets, we included 12 free parameters
representing offsets relative to the \cite{Roemer:1965} velocities,
which are the most numerous and have the longest baseline.  The MCMC
chains had 10,000 links after burn-in, and the priors adopted for the
adjustable parameters were all uniform over suitable ranges.
Convergence was checked by inspecting the chains visually, and by
requiring a Gelman-Rubin statistic of 1.05 or smaller
\citep{Gelman:1992}.

Initial solutions indicated many of the datasets had velocity
uncertainties that are underestimated, based on their $\chi^2$
values. In two cases (\citealt{Dinshaw:1989}, and the Usenko
velocities) they appeared to be significantly overestimated. We first
scaled down the errors for those two groups arbitrarily by a factor of
5, and then allowed for an additional source of uncertainty in each of
the 13 datasets by solving for separate ``jitter'' values, added
quadratically to the formal uncertainties. This extra jitter may be
caused in part by unstable zeropoints within each dataset, by
imperfect modelling and subtraction of the
pulsation variation, or by other unrecognised periodicities.

\begin{figure*}
\epsscale{1.17}
\plotone{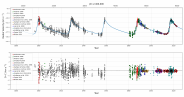}

\figcaption{Radial velocities of Polaris disaffected from the
  pulsation variations, and adjusted for the offsets given in
  Table~\ref{tab:offsets}. The ``AF1983+K1996" dataset 
  combines the \cite{ArellanoFerro:1983a} velocities
  adjusted by +0.5~\kms, and those of \cite{Kamper:1996}.
  The curve is our model from
  Table~\ref{tab:mcmc}. Compare with Figure~\ref{fig:raw.rvs}.
  Velocity residuals are shown at the bottom on an expanded scale. \label{fig:final.rvs}}

\end{figure*}

\begin{figure}
\epsscale{1.17}
\plotone{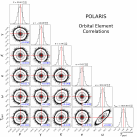}

\figcaption{Corner plot for the standard orbital elements of Polaris,
showing the correlations. The contours correspond to confidence levels
of 1, 2, and 3$\sigma$, and the values in each panel are the
correlation coefficients.\label{fig:correlations}}

\end{figure}

The resulting orbital elements are presented in Table~\ref{tab:mcmc},
and the velocity offsets and jitter terms are given separately in
Table~\ref{tab:offsets}. A plot of the velocities and our model is
seen in Figure~\ref{fig:final.rvs}, in which all offsets listed in
Table~\ref{tab:offsets} have been applied. The scale is the same as in
Figure~\ref{fig:raw.rvs}, and shows the reduction in the scatter after
the removal of the pulsation variation. Correlations among the
standard orbital elements are shown in Figure~\ref{fig:correlations}.

\setlength{\tabcolsep}{6pt}
\begin{deluxetable}{lc}
\tablecaption{Updated Spectroscopic Orbital Elements for Polaris \label{tab:mcmc}}
\tablehead{
\colhead{~~~~~~~~Parameter~~~~~~~~} &
\colhead{Value}
}
\startdata
$P$ (yr)               & $29.4330 \pm 0.0079$\phn  \\
$\gamma$ (\kms)        & $-16.084 \pm 0.025$\phn\phs   \\
$K$ (\kms)             & $3.7409 \pm 0.0075$   \\
$\sqrt{e}\cos\omega$   & $+0.4175 \pm 0.0040$\phs  \\
$\sqrt{e}\sin\omega$   & $-0.6672 \pm 0.0026$\phs  \\
$T_{\rm peri}$ (yr)    & $2016.801 \pm 0.011$\phm{222}  \\ [1ex]
\hline \\ [-1.5ex]
\multicolumn{2}{c}{Derived properties} \\ [1ex]
\hline \\ [-1.5ex]
$e$                    & $0.6195 \pm 0.0015$   \\
$\omega$ (degree)        & $302.04 \pm 0.34$\phn\phn     \\
$M_{\rm Ab} \sin i / (M_{\rm Aa} + M_{\rm Ab})^{2/3}$ ($M_{\sun}$) & $0.30442 \pm
0.00075$ \\
$a_{\rm Aa} \sin i$ ($10^6$ km) & $434.1 \pm 1.1$\phn\phn 
\enddata
\tablecomments{The values listed correspond to the mode of
the posterior distributions, with their 68.3\% credible intervals,
which are symmetrical in all cases.
The centre-of-mass velocity $\gamma$ uses the
  \cite{Roemer:1965} dataset as the reference. The last two
  lines give the coefficient of the minimum secondary mass,
  and the linear semimajor axis of the primary.}
\end{deluxetable}
\setlength{\tabcolsep}{6pt}

\setlength{\tabcolsep}{4pt}
\begin{deluxetable}{lcc}
\tablecaption{Radial Velocity Offsets and Jitter Values \label{tab:offsets}}
\tablehead{
\colhead{Source} &
\colhead{RV Offset} &
\colhead{RV Jitter} 
\\
\colhead{} &
\colhead{(\kms)} &
\colhead{(\kms)} 
}
\startdata
\phn(1)  \cite{Roemer:1965}\tablenotemark{a}          & \nodata                      & $0.681^{+0.022}_{-0.022}$  \\ [1ex]
\phn(2)  \cite{Hartmann:1901}                         & $+1.39^{+0.19}_{-0.18}$      & $0.51^{+0.24}_{-0.26}$     \\ [1ex]
\phn(3)  \cite{Kustner:1908}                          & $+1.05 ^{+0.51}_{-0.52}$     & $0.75^{+0.80}_{-0.45}$     \\ [1ex]
\phn(4)  \cite{ArellanoFerro:1983a}\tablenotemark{b}  & $+1.027^{+0.043}_{-0.043}$   & $0.524^{+0.032}_{-0.029}$  \\ [1ex]
\phn(5)  \cite{Dinshaw:1989}                          & $-11.829^{+0.050}_{-0.051}$  & $0.469^{+0.031}_{-0.026}$  \\ [1ex]
\phn(6)  \cite{Gorynya:1992}\tablenotemark{c}         & $+1.488^{+0.069}_{-0.070}$      & $0.500^{+0.053}_{-0.045}$  \\ [1ex]
\phn(7)  \cite{Hatzes:2000}                           & $-14.117^{+0.030}_{-0.030}$  & $0.027^{+0.019}_{-0.016}$  \\ [1ex]
\phn(8)  \cite{Usenko:2015}\tablenotemark{d}          & $+0.068^{+0.042}_{-0.042}$   & $0.387^{+0.028}_{-0.026}$  \\ [1ex]
\phn(9)  \cite{Kamper:1998}                           & $-0.222^{+0.028}_{-0.028}$   & $0.1061^{+0.0070}_{-0.0060}$  \\ [1ex]
(10)  \cite{Eaton:2020}                            & $+1.120^{+0.027}_{-0.026}$   & $0.0444^{+0.0078}_{-0.0110}$  \\ [1ex]
(11)  \cite{Lee:2008}                              & $-17.794^{+0.027}_{-0.028}$  & $0.1103^{+0.0054}_{-0.0046}$  \\ [1ex]
(12)  \cite{Bucke:2021}                            & $-18.387^{+0.027}_{-0.027}$  & $0.0976^{+0.0110}_{-0.0094}$     \\ [1ex]
(13)  \cite{Anderson:2019}                         & $+0.551^{+0.027}_{-0.027}$   & $0.0977^{+0.0064}_{-0.0054}$  
\enddata
\tablecomments{The values listed correspond to the mode of
the posterior distributions, with their 68.3\% credible intervals.
The offsets are to be added to each dataset to place the RVs
on the frame of the \cite{Roemer:1965} velocities.
The RV jitter values are added quadratically
to the internal errors. Sources are numbered as in Table~\ref{tab:rvs}.}
\tablenotetext{a}{Reference dataset for all RV offsets.}
\tablenotetext{b}{This dataset includes the velocities of
  \cite{Kamper:1996}.}
\tablenotetext{c}{This dataset includes the velocities of
\cite{Gorynya:1998}.}
\tablenotetext{d}{This dataset includes the velocities of
  \cite{Usenko:2016, Usenko:2017, Usenko:2018, Usenko:2020}.}
\end{deluxetable}
\setlength{\tabcolsep}{6pt}

%%%%%%%%%%%%%%%%%%%%%%%%%%%%%%%%%%%%%%%%%%%%%%%%%%%%%%%%%%%%%%%%%%%%%
\subsection{Period Variations and the $O\!-\!C$ Diagram}
\label{sec:o-c}
%%%%%%%%%%%%%%%%%%%%%%%%%%%%%%%%%%%%%%%%%%%%%%%%%%%%%%%%%%%%%%%%%%%%%

Changes in the period of pulsation of Polaris have long been known
from the work of visual observers \citep[e.g.,][]{Stebbins:1946}, and
were shown clearly also in the extensive RV measurements by
\cite{Roemer:1965} (see her Figure~46). Most observers have concluded
that the period has been increasing at a fairly constant rate between
3 and 5~s~yr$^{-1}$, based on a quadratic fit to the $O\!-\!C$
diagram.  Some authors have drawn attention to deviations from that
shape, and have suggested 5th order or even 7th order approximations
\citep{Fernie:1993, Kamper:1998}. In a comprehensive review of the
existing material for Polaris, \cite{Turner:2005} remeasured the
timings from all available visual, photometric, and RV observations
going back to 1844, and up to 2004. They concluded that the period has
been monotonically increasing at a rate of approximately
4.5~s~yr$^{-1}$, but with a brief interruption between about 1963 and
1966, during which it appeared to have suddenly decreased. The period
then resumed its increase at about the same rate as before. Other
authors have followed \cite{Turner:2005} in showing separate parabolic
fits to the timings in the $O\!-\!C$ diagram before and after 1965,
and have reported that new measurements until 2011 still appear to fit
the more recent trend fairly well \citep[e.g.,][]{Neilson:2012}.

\begin{figure}
\epsscale{1.17}
\plotone{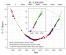}
\figcaption{$O\!-\!C$ diagram for Polaris based on visual, photoelectric,
  and spectroscopic timing measurements by \cite{Turner:2005}. New
  spectroscopic values after 2005 are added from the present work. The
  curves are weighted quadratic fits to the \cite{Turner:2005} data
  before and after 1963 (excluding our own), implying fairly
  consistent period increases of $dP/dt = 4.45 \pm 0.04$ and $5.07 \pm
  0.57$~s~yr$^{-1}$, respectively. The new spectroscopic data show an obvious
  departure from the fit after about 2012. \label{fig:omc}}

\end{figure}

Figure~\ref{fig:omc} reproduces the photometric and spectroscopic
$O\!-\!C$ measurements from the work of \cite{Turner:2005}, to which
we have added the subset of our own spectroscopic values from
Table~\ref{tab:omc} after 2005. The curves displayed are our weighted
quadratic fits to the \cite{Turner:2005} measurements, before and
after 1965. The most recent of our data after about 2012 show a clear
departure from the quadratic fit, implying the period has stopped
increasing and is now becoming shorter. Early indications of this were
discussed by \cite{Spreckley:2008}, based on photometric observations
collected by the Solar Mass Ejection Imager (SMEI) instrument on board
the Coriolis satellite, between 2003 and 2007.

\begin{figure}
\epsscale{1.16}
\plottwo{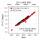}{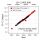}
\figcaption{Determination of the optimal shift to be applied to
our measured times of minimum RV in order to align them with
the times of maximum brightness from the SMEI observations \citep{Spreckley:2008}.
The $O\!-\!C$ diagram on the left uses the offset adopted by \cite{Turner:2005}
to adjust the RV timings.
The one on the right uses the value of $-0.21$~d advocated by
\cite{ArellanoFerro:1983a}, and clearly works best for our data.
\label{fig:omcoffset}}
\end{figure}

We used an updated version of the $O\!-\!C$ diagram to infer the
instantaneous values of the pulsation period as a function of time.
The data for this diagram included all our newly derived spectroscopic
timings from Table~\ref{tab:omc}, the photometric measurements of
\cite{Turner:2005}, and the highly precise times of maximum light from
the SMEI instrument by \cite{Spreckley:2008}. Those timings are
contemporaneous with some of our spectroscopic determinations, and
support a value for the offset between the times of velocity minima
and light maxima of $-0.21$~d (see Figure~\ref{fig:omcoffset}), as
found by \cite{ArellanoFerro:1983a}. We applied this value to all our
spectroscopic timings to align them with the photometric values. We
have also added 3 photometric timings by \cite{Bruntt:2008} from the
WIRE satellite, and 5 photometric timings by \cite{Neilson:2012},
extracted by digitising their Figure~1. In addition, Polaris has been
observed by the Transiting Exoplanet Survey Satellite
\citep[TESS;][]{Ricker:2015} in several sectors (\#19, 20, 25, 26, 40,
47, 52, 53, and 60, as of this writing). We downloaded the photometry
from the Mikulski Archive for Space Telescopes
(MAST)\footnote{\url{https://archive.stsci.edu/}}, and measured the
times of maximum light by fitting a sine function to the top half of
each cycle.\footnote{Because Polaris is such a bright star, it
  saturates the TESS detectors and causes ``bleed trails" extending
  many CCD rows above and below the saturated pixels. This requires
  very large photometric apertures in order to capture all the flux
  (see
  \url{https://heasarc.gsfc.nasa.gov/docs/tess/observing-technical.html}),
  and makes it difficult to extract precise photometry from these
  observations, particularly as the extent of the bleed trails may
  vary along the pulsation cycle.  While we do not expect this to bias
  our determination of the times of maximum brightness, it may
  compromise the determination of light amplitudes, and for this
  reason we do not use the TESS photometry in the next section for
  that purpose.}  These 52 new measurements are listed in
Table~\ref{tab:tess}.

\setlength{\tabcolsep}{8pt}
\begin{deluxetable}{cccc}
\tablecaption{New Times of Maximum Light from TESS \label{tab:tess}}
\tablehead{
\colhead{BJD} &
\colhead{$\Delta t$} &
\colhead{$E$} &
\colhead{$O-C$}
\\
\colhead{(2,400,000+)} &
\colhead{(day)} &
\colhead{} &
\colhead{(day)}
}
\startdata
  58817.473  &  $-0.011$  & 7695  &  $13.349 \pm 0.025$ \\
  58821.472  &  $-0.011$  & 7696  &  $13.378 \pm 0.025$ \\
  58825.434  &  $-0.011$  & 7697  &  $13.371 \pm 0.025$ \\
  58829.255  &  $-0.011$  & 7698  &  $13.223 \pm 0.025$ \\
  58833.365  &  $-0.011$  & 7699  &  $13.364 \pm 0.025$ 
\enddata
\tablecomments{The $\Delta t$ values in column 2 are the additive light
travel time corrections to reduce the measured times of maximum light
in column 1 to the barycentre of the 30~yr binary. Cycle numbers $E$
are counted from the reference epoch given in eq.[\ref{eq:ephemeris}].
(This table is available in its entirety in machine-readable form).}
\end{deluxetable}

\begin{figure}
\epsscale{1.15}
\plotone{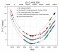}
\plotone{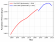}
\figcaption{\emph{Top}: Updated $O\!-\!C$ diagram for Polaris (grey
  crosses) that includes the visual and photoelectric timing
  measurements by \cite{Turner:2005}, our own spectroscopically
  determined timings from Table~\ref{tab:omc}, the SMEI timings from
  \cite{Spreckley:2008}, 3 additional measurements by
  \cite{Bruntt:2008} from the WIRE satellite, and 5 others by
  \cite{Neilson:2012}. The solid curves pre- and post-1965 are spline
  fits to the data. The photometric and spectroscopic timings are also
  shown separately to illustrate their time coverage, by plotting them
  in parallel sequences (dashed lines) offset vertically from the
  central sequence (crosses) for clarity. \emph{Bottom}: The
  time-dependent instantaneous pulsation period derived from the $O\!-\!C$
  diagram, before and after the interruption in 1965.\label{fig:period}}

\end{figure}

All of the additional times of maximum light mentioned above were
corrected for light travel time, for consistency.  The updated
$O\!-\!C$ diagram is shown in the top panel of
Figure~\ref{fig:period}, along with spline fits to the data before and
after the 1965 break. Using these spline fits, we derived a smooth
representation of the evolution of the instantaneous pulsation period
over the last 175~yr, which is shown in the lower panel.  A turnover
is seen to have occurred around 2010, and the period is now becoming
shorter.  The sudden drop in 1965 is close to a minute, and the total
change between 1845 and the maximum around 2010 is about 12 minutes,
corresponding to an average of $\sim$4.5~s~yr$^{-1}$, as found
previously.

The sharpness and amplitude of the decline in the mid 1960s reported
by \cite{Turner:2005} rely strongly on a single measurement of a time
of maximum brightness from late 1962 (HJD~2,437,971.594).  This
determination is based on 3 narrow-band photometric observations of
Polaris by \cite{Williams:1966}, transformed to the visual band, of
which two were made on consecutive nights and the other 77 days later
(19 pulsation cycles). Regrettably, no other useful photometric or RV
measurements during this time have been identified in the literature,
although it is possible they exist.

%%%%%%%%%%%%%%%%%%%%%%%%%%%%%%%%%%%%%%%%%%%%%%%%%%%%%%%%%%%%%%%%%%%%%
\subsection{Pulsation Amplitude Variations}
\label{sec:amplitude}
%%%%%%%%%%%%%%%%%%%%%%%%%%%%%%%%%%%%%%%%%%%%%%%%%%%%%%%%%%%%%%%%%%%%%

The oscillations of Polaris as measured photometrically or
spectroscopically have the same shape \citep[e.g.,][]{Klagyivik:2009},
and differ only by a phase shift, as noted in the previous section,
and by a scale factor.  Historical estimates of the pulsation
amplitude based on brightness measurements have generally been more
problematic to interpret. They are typically less precise (made
visually for the oldest estimates), less homogeneous, and it is well
known for Cepheids and similar variable stars that the brightness
changes are smaller in the visible than in blue light \citep[see,
  e.g.,][]{Stebbins:1946, Klagyivik:2009}.  This makes comparisons
difficult when the wavelength of the observation is not well known,
leading to increased scatter \citep[e.g.,][]{ArellanoFerro:1983a}. We
return to this below.

Measures of the radial velocity amplitude of the pulsations in Polaris
are less ambiguous, but are not without their challenges. Because of
the complex velocity fields in the atmospheres of Cepheids, in
principle the measured velocity range can depend to some extent on the
depth of formation of the spectral lines used to measure the RVs
\citep{Sasselov:1990}, as well as on the measuring technique
\cite[see, e.g.,][and references therein]{Anderson:2018b,
  Anderson:2019}. These effects on the amplitude do not seem to be as
overwhelming as in the case of brightness measurements, but should
nevertheless be kept in mind.  For the RV sources from the literature
used here, the approximate wavelength ranges of the observations are
listed in Table~\ref{tab:historical.rvs}, when known.

\begin{figure}
\epsscale{1.15}
\plotone{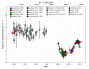}
\plotone{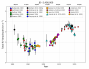}

\figcaption{\emph{Top:} Pulsation amplitude of Polaris (peak-to-peak)
  measured in this work from all available RV observations
  (Table~\ref{tab:omc}). One additional measurement has been included,
  from the work of \cite{Fagas:2009} (see the Appendix).
  \emph{Bottom:} Enlargement showing the last
  four decades of amplitude measurements in closer
  detail.\label{fig:amp}}

\end{figure}

RV amplitude changes were first examined by \cite{Roemer:1965}, who
concluded there was no obvious variation during the 60+~yr of Lick
Observatory observations (1896--1958). \cite{Turner:2005} added
another 40~yr of amplitude measurements from other RV sources. Their
Figure~6 displaying all spectroscopic and photometric measurements
together shows considerable dispersion, but this is largely due to the
photometric estimates, which are typically more uncertain.
Figure~\ref{fig:amp} (top) is a graphical representation of our RV
amplitude measurements from Table~\ref{tab:omc}, covering 125~yr.  As
is customary in the Cepheid literature, we plot the full amplitude of
the variation ($2A_{\rm puls}$), rather than the semiamplitude.

It appears to us that the Lick observations do suggest a very slight
reduction in the amplitude up to 1958, in agreement with
\cite{Turner:2005}, whereas \cite{Roemer:1965} chose to be more
conservative.  As mentioned earlier, Polaris seems to have been
neglected by spectroscopists for the following two decades, during
which there happened to be a sharp reduction in the pulsation
amplitude.  After 1980, the velocities from at least half a dozen
independent observers showed the tail end of this amplitude decline,
reaching a minimum peak-to-peak value of only $\sim$1.5~\kms. The
amplitude then levelled off around 1990, and stayed low until about
1998. Beginning in 2002 it began to increase again, reaching a maximum
of 3.5~\kms\ around 2015. The lower panel of Figure~\ref{fig:amp}
zooms in on the last 40~yr of data.  The three sources of measurements
during the last decade (\citealt{Anderson:2019}, \citealt{Bucke:2021},
and Usenko and collaborators) display increased scatter. The
measurements from the first two of these studies show little change in
the amplitude after 2015, although they are at slightly different
levels, whereas those of Usenko suggest another rather steep decline
\citep{Usenko:2020}.  The reasons for these discrepancies are
unclear. We note also that the amplitudes from the \cite{Bucke:2021}
data (2005--2022) appear systematically lower by 0.3 or 0.4~\kms\ than
other estimates obtained near in time. Finally, some of the scatter in
Figure~\ref{fig:amp} may be real, as the RV range of the pulsations
from cycle to cycle could be variable to some degree.  And as pointed
out earlier, differences in the spectral lines used, or in the
techniques to measure the RVs, can also add scatter.

\begin{figure}
\epsscale{1.15}
\plotone{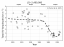}
\figcaption{Photometric $V$-band estimates of the pulsation amplitude
  of Polaris by \cite{Turner:2005}, scaled up by a factor of 50 to
  match the peak-to-peak amplitude measurements from the RVs. A
  dashed line has been drawn by eye to mark the general trend,
  which is consistent with that shown in Figure~\ref{fig:amp} (top).
  \label{fig:ampturner}}

\end{figure}

While there is a gap in the spectroscopic measures of the pulsation
amplitude during the critical period between 1960 and 1980, a few
photometric estimates do exist in this interval, although they exhibit
considerable scatter, particularly before 1980. In
Figure~\ref{fig:ampturner} we represent all light amplitude
determinations by \cite{Turner:2005}, which the author measured in a
uniform way and kindly provided to us. They are all presumed to
correspond to the $V$ band, and we have scaled them here so as to
match the velocity amplitudes ($\Delta RV = 2A_{\rm puls}$) using
$\Delta RV = 50\,\Delta V$, where the coefficient of
50~\kms~mag$^{-1}$ is the one adopted by \cite{Turner:2005}.  The
decline in the light amplitude during 1960--1980 is fairly obvious,
and an approximate representation is shown by the dashed line drawn by
eye. It is consistent with the trend observed in Figure~\ref{fig:amp}
(top).

\begin{figure}
\epsscale{1.15}
\plotone{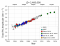}
\figcaption{Comparison between the total pulsation amplitudes of
  Polaris from RVs and those based on the 2003--2011 SMEI photometry,
  as measured by \cite{Anderson:2019} and kindly provided by the
  author.  The SMEI amplitudes have been scaled up here by a factor of 77
  to match the peak-to-peak velocity amplitudes. This coefficient is
  different than in the visible because of the redder filter response
  of the photometric observations (roughly Johnson $R$), but is
  similar to that found by \cite{Bruntt:2008}. The near perfect
  match of the slopes supports the accuracy of the photometric
  amplitudes. Three additional photometric amplitude measurements from
  the WIRE satellite are also shown. They were extracted from Figure~4
  by \cite{Bruntt:2008}. \label{fig:smei}}

\end{figure}

Much more precise measures of the peak-to-peak brightness amplitude
have been obtained by several authors from the 2003--2011 SMEI
observations \citep{Bruntt:2008, Svanda:2017, Anderson:2019}. All
three studies showed that the pulsation amplitude more than doubled
during that period.\footnote{At the time of the \cite{Bruntt:2008}
  paper, only about half of the SMEI data were available, but the
  amplitudes showed the same trend as the full dataset later analysed
  by others.}  However, \cite{Anderson:2019} drew attention to a
strong correlation between the growing photometric amplitude and the
mean magnitude of Polaris measured from the same observations (his
Figure~7). The correlation is in the sense that Polaris seemed to be
getting fainter with time, by roughly 0.08~mag over that period. As
this is unexpected, Anderson cautioned that the apparent dimming could be
of instrumental origin, such as from a change in the detector's
non-linearity properties, and that this might also have resulted in a
spurious change in the light amplitude.  Radial velocity measurements
are now available over the same period as the SMEI observations, and
confirm, as mentioned earlier, that the pulsation amplitude was in
fact increasing. We illustrate this in Figure~\ref{fig:smei}, which
compares the SMEI amplitudes from \cite{Anderson:2019} against those
from the highest quality spectroscopic datasets over those years
\citep{Lee:2008, Fagas:2009, Anderson:2019, Eaton:2020}. The agreement
between these two completely independent ways of measuring the
pulsation amplitude is excellent. In addition to supporting the
accuracy of the SMEI amplitude measurements, this would also seem to
lend credence to the finding by \cite{Anderson:2019} that the average
brightness of Polaris was decreasing at the time (but see below).

Accurate measurements of the mean magnitude of Polaris over long
periods of time are challenging, to say the least, and in that sense
the SMEI observations are unique for their precision and continuity.
We are not aware of any detailed study of the long-term evolution of
its brightness, other than brief reports by \cite{Engle:2004a,
  Engle:2004b, Engle:2014}.  These authors claimed that Polaris seems
to have become brighter in the last century by one or two tenths of a
magnitude, and that examination of historical records going back to
Ptolemy and Hipparchus suggest it may have been much fainter by a
magnitude or more some 2000 years ago.\footnote{See also the popular
  account in Sky \& Telescope, Vol.\ 137, No.\ 3, p.\ 14 (March
  2019).}  We note, however, that \cite{Neuhauser:2022} have expressed
some reservations about the significance of the differences between
the oldest measurements and current ones. \cite{Turner:2009} also
doubted those claims, and concluded that the brightness of Polaris
appears to have been constant at $V \approx 2.00$ for the past century
and a half.

%%%%%%%%%%%%%%%%%%%%%%%%%%%%%%%%%%%%%%%%%%%%%%%%%%%%%%%%%%%%%%%%%%%%%
\section{Discussion}
\label{sec:discussion}
%%%%%%%%%%%%%%%%%%%%%%%%%%%%%%%%%%%%%%%%%%%%%%%%%%%%%%%%%%%%%%%%%%%%%

The spectroscopic orbit of Polaris is a key ingredient for determining
the dynamical masses of the components, as it provides the so-called
mass function of the binary, or equivalently, the minimum secondary
mass $M_{\rm Ab} \sin i = (P/2\pi G)^{1/3} \sqrt{1-e^2} K (M_{\rm
  Aa}+M_{\rm Ab})^{2/3}$.  In this expression, the mass sum on the
right is obtained from the astrometric orbit and the parallax, through
Kepler's third law.  The precision in the masses is currently limited
by the astrometry, which has very sparse phase coverage from the five
available HST observations, and in some cases sizeable uncertainties
due to the difficulty of the measurements \citep{Evans:2008,
  Evans:2018}. Nevertheless, there is also some contribution from the
uncertainty in the spectroscopic orbital elements.
Table~\ref{tab:orbits} compares our improved spectroscopic elements to
others published previously. In earlier solutions, the contribution of
the spectroscopy to the uncertainty in the masses was at the level of
1--3\%, in most cases. By using twice the number of RV measurements
now covering 4.3 cycles of the binary, the present work reduces this
contribution by an order of magnitude. The dominant term in the mass
error budget will then continue to come from the astrometry,
particularly the inclination angle and the semimajor axis, until
additional coverage of the orbit can be obtained.

\setlength{\tabcolsep}{3pt}
\begin{deluxetable*}{lcccccc}
\tablecaption{Spectroscopic Orbital Solutions for Polaris \label{tab:orbits}}
\tablehead{
\colhead{Source} &
\colhead{$P$} &
\colhead{$\gamma$} &
\colhead{$K$} &
\colhead{$e$} &
\colhead{$\omega$} &
\colhead{$T_{\rm peri}$}
\\
\colhead{} &
\colhead{(yr)} &
\colhead{(\kms)} &
\colhead{(\kms)} &
\colhead{} &
\colhead{(degree)} &
\colhead{(yr)}
}
\startdata
%                                    P                           gamma                          K                        ecc                     w1                             Tperi
\cite{Moore:1929}       &   29.6\phn                      &  $-17.4$\phn                      &  4.05                 &  0.63                 &  332.0\phn                      &  1899.5\phn\phn                        \\
\cite{Gerasimovic:1936} &   29.6 (assumed)            &  $-16.5$\phn                      &  4.35                 &  0.50                 &  315.8\phn                      &  1899.8\phn\phn                        \\
\cite{Roemer:1965}      &   $30.46 \pm 0.15$\phn      &  $-16.41$\phn                     &  $4.09 \pm 0.15$      &  $0.639 \pm 0.018$    &  $307.2 \pm 2.7$\phn\phn    &  $1928.48 \pm 0.12$\phm{222}   \\
\cite{Kamper:1984}      &   $30.03 \pm 0.07$\phn      &  $-16.31 \pm 0.04$\phn\phs    &  $4.01 \pm 0.06$      &  $0.655 \pm 0.013$    &  $307.2 \pm 1.1$\phn\phn    &  $1929.86 \pm 0.11$\phm{222}   \\
\cite{Kamper:1996}      &   $29.59 \pm 0.02$\phn      &  $-16.42 \pm 0.03$\phn\phs    &  $3.72 \pm 0.03$      &  $0.608 \pm 0.005$    &  $303.01 \pm 0.75$\phn\phn  &  $1928.48 \pm 0.08$\phm{222}   \\
\cite{Turner:2009}      &   $29.71 \pm 0.09$\phn      &  $-15.90 \pm 0.06$\phn\phs    &  $4.41 \pm 0.07$      &  $0.543 \pm 0.010$    &  $309.6 \pm 0.7$\phn\phn    &  $1928.57 \pm 0.06$\phm{222}   \\
\cite{Turner:2009}      &   $29.80 \pm 0.05$\phn      &  $-15.87 \pm 0.05$\phn\phs    &  $4.23 \pm 0.07$      &  $0.52 \pm 0.01$      &  $301 \pm 2$\phn\phn        &  $1928.50 \pm 0.12$\phm{222}   \\
\cite{Anderson:2019}    &   $29.32 \pm 0.11$\phn      &  $-15.387 \pm 0.040$\phn\phs  &  $3.768 \pm 0.073$    &  $0.620 \pm 0.008$    &  $307.2 \pm 2.5$\phn\phn    &  $2016.91 \pm 0.10$\phm{222}   \\
\cite{Usenko:2020}      &   $29.25 \pm 0.03$\phn      &  $-16.61 \pm 0.12$\phn\phs    &  $3.93 \pm 0.12$      &  $0.633 \pm 0.044$    &  $302.5 \pm 2.7$\phn\phn    &  $1987.22 \pm 0.10$\phm{222}   \\
\cite{Bucke:2021}       &   $29.31 \pm 0.07$\phn      &  $-17.1 \pm 0.4$\phn\phs      &  $3.74 \pm 0.06$      &  $0.574 \pm 0.010$    &  $300.7 \pm 3.0$\phn\phn    &  $2016.77 \pm 0.10$\phm{222}   \\
This work               &   $29.4330 \pm 0.0079$\phn  &  $-16.084 \pm 0.025$\phn\phs  &  $3.7409 \pm 0.0075$  &  $0.6195 \pm 0.0015$  &  $302.04 \pm 0.34$\phn\phn  &  $2016.801 \pm 0.011$\phm{222} 
\enddata
\end{deluxetable*}
\setlength{\tabcolsep}{6pt}

The most recent dynamical mass determination for Polaris by
\cite{Evans:2018}, giving $3.45 \pm 0.75~M_{\sun}$ and $1.63 \pm
0.49~M_{\sun}$ for the primary and secondary, respectively, relied on
spectroscopic elements from previous work, which were held fixed.  The
elements $P$, $e$, and $\omega$ were adopted from \cite{Kamper:1996},
while the time of periastron passage $T_{\rm peri}$ was taken from
\cite{Wielen:2000}.  Our improved results show subtle differences with
those elements, particularly in the period and eccentricity
(Table~\ref{tab:orbits}), which may have a non-negligible impact on
the masses because the astrometric measurements only cover a small
fraction of the orbit so far.  However, given that the mass errors
remain dominated by the astrometry, there is relatively little reward
in repeating that determination at the present time.  Further
astrometric observations of Polaris are currently underway
\citep[see][]{Evans:2018}, and may soon lead to significant
improvements.

Up until the mid 1990s, there were differing opinions about the mode
in which Polaris pulsates, some arguing for the fundamental mode, and
others for the first or even second overtone. See, e.g.,
\cite{Evans:2018}, \cite{Bond:2018}, \cite{Engle:2018}, or
\cite{Anderson:2018a} for summaries of those discussions.  Much of the
debate hinged on the uncertain distance of Polaris (and hence its
radius and luminosity). A variety of methods have been used for that
purpose, including trigonometric parallaxes from the ground
\citep{vanAltena:1995} and with HST \citep{Bond:2018}, photometric
parallaxes of Polaris~B \citep{Turner:1977, Turner:2013}, absolute
magnitude estimates based on spectroscopic line ratios calibrated
against supergiants with well-established luminosities
\citep{Turner:2013}, and others.  The distance issue was eventually
settled by the Hipparcos mission, although not without further debate
\citep[see][]{Turner:2013, vanLeeuwen:2013, Bond:2018}. The Gaia
mission finally confirmed the accuracy of the Hipparcos result by
measuring the parallax of Polaris~B, and increased the precision by
about a factor of ten. Polaris is now considered to be a first
overtone pulsator \citep[e.g.,][]{Anderson:2018a, Anderson:2019},
consistent with its small amplitude, symmetric light curve, and its
rapid rate of period change, which in this type of objects is about an
order of magnitude faster than seen in Cepheids that are pulsating in
the fundamental mode \citep[e.g.,][]{Szabados:1983}.  Secular period
changes in Cepheids are not uncommon \citep[see,
  e.g.,][]{Csornyei:2022}. A period increase is indicative of
evolution toward cooler temperatures, and for fundamental mode
pulsators, the rates are broadly in agreement with predictions from
stellar evolution models.

Except for a brief interruption in the mid 1960s, the monotonic
increase in the pulsation period of Polaris until about 2005, at a
rapid rate of $\sim$4.5~s~yr$^{-1}$, was considered a sign that the
star was evolving to the right in the H-R diagram
\citep[e.g.,][]{Turner:2006}, and was about to exit the instability
strip. This idea was driven in part by the declining pulsation
amplitude, and by the belief that Polaris was near the red edge of the
instability strip, whereas revisions in the distance locate it closer
to the blue edge \citep[see, e.g.,][]{Anderson:2019, Ripepi:2021}.
The fact that the period is now becoming shorter also complicates this
picture, as evolution is expected to proceed only in one
direction. The glitch circa 1965 has defied explanation as well,
although \cite{Turner:2005} have speculated that it could be accounted
for by a sudden increase in the mass of the Cepheid, such as by the
ingestion of a large ($\sim$7~$M_{\rm Jup}$) planet.
\cite{Evans:2002, Evans:2018} have argued that for overtone pulsators
like Polaris, period changes are not necessarily caused by evolution
across the instability strip.  Instead, they may simply reflect more
erratic behaviour related to the pulsation itself, which is still
poorly understood.  The irregular fluctuations in the pulsation
amplitude would seem to support that notion.

It has been proposed that some of these irregular changes in the
properties of the pulsation may be related to the binarity of Polaris
\citep[e.g.,][]{Dinshaw:1989, Usenko:2020}. Discontinuities in the
$O\!-\!C$ diagram, such as the one in the mid 1960s, have long been
seen also in other Cepheids \citep[e.g.,][]{Szabados:1989,
  Szabados:1991, Csornyei:2022}, and are sometimes referred to as
phase jumps or phase slips. \cite{Szabados:1992} pointed out that such
features appear to be common in binary Cepheids, more so than in
Cepheids that are single, and that they seem to repeat with a cycle length
similar to the period of the binary, or multiples of it. They would be
explained by the perturbing influence of the companion on the upper
layers of the atmosphere of the Cepheid, where the pulsations take
place, during or near the times of periastron passage.

The separation between Polaris and its 30~yr companion at periastron
is approximately 6.2~au. This follows from the angular semimajor axis
of the astrometric orbit by \cite{Evans:2018} ($a = 0\farcs12$), our
value for the eccentricity ($e = 0.62$), and the parallax of Polaris~B
from the Gaia mission. The latest Gaia data release
\citep[DR3;][]{Gaia:2022} puts the parallax at $\pi_{\rm Gaia} =
7.305$~mas, including a zeropoint correction by
\cite{Lindegren:2021}. Polaris has a measured angular diameter of
$3.123 \pm 0.008$~mas \citep{Merand:2006}, which corresponds to a
physical radius of 46~$R_{\sun}$ at the distance from Gaia
(137~pc). The periastron separation is therefore about 29 times the
radius of the Cepheid.

\begin{figure}
\epsscale{1.15}
\plotone{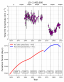}
\figcaption{RV amplitude of Polaris (top) and instantaneous pulsation period
(bottom) as a function of time (from Figures~\ref{fig:period} and
\ref{fig:amp}), with periastron
passages marked with dashed lines and labelled.\label{fig:periastron}}

\end{figure}

In Figure~\ref{fig:periastron} we reproduce the pulsation velocity
amplitudes and instantaneous pulsation period of Polaris from earlier
figures. The dashed lines mark the periastron passages according to
our updated spectroscopic orbit.  There are intriguing coincidences
between features in these diagrams and the dates of minimum separation
of the binary. In the top panel, the 1957 periastron came just before
the sharp drop in the RV amplitude that took place between 1960 and
1980. The exact date at which this decline began is difficult to
pinpoint due to a lack of measurements.  It may have started a few
years into the 1960--1980 gap, as Figure~\ref{fig:ampturner} might
suggest.  It would not be surprising if the period jump in the lower
panel coincided with that instant, as it also occurred just a few
years after closest approach in 1957.  The following periastron
passage, in 1987, came just before the RV amplitudes reached their
lowest point and reversed course. At about the same time, there is a
slight change in the slope of the curve representing the evolution of
the pulsation period, although the reality of this feature is unclear
as it depends to some extent on the smoothing applied to the
curve. The most recent periastron passage of 2016 occurred just as the
RV amplitudes appeared to reach a peak. At about this time, or
slightly earlier, the period of Polaris began to decrease.

Earlier periastron passages appear less remarkable, as the data are
either sparser or of lesser quality. There is a hint that the 1928
passage may coincide with a downward kink in the distribution of
velocity amplitudes, although we cannot be certain given the precision
of the data.  The passage of 1899 occurred near an inflection point in
the period diagram, but again, the data here are scarce, particularly
before 1900 (Figure~\ref{fig:period}, top).

Taken together, these instances are highly suggestive of a
cause-and-effect connection between changes in the pulsation behaviour
of Polaris and the recurring approaches of the companion every
30~yr. The implication would be that the upper atmosphere of the
Cepheid is undergoing enhanced, forced tidal perturbations induced by
the secondary.  Such tidal forcing in eccentric binaries is also seen
in a very different context, in the class of main-sequence objects
known as ``heartbeat stars" \citep[see, e.g.,][]{Welsh:2011}. In these
systems, which are typically very eccentric, each periastron passage
is responsible for exciting a rich set of stellar pulsations with a
large number of modes, giving the lightcurves their characteristic
appearance of an electrocardiogram.

In overtone Cepheid pulsators such as Polaris, \cite{Evans:2002} have
argued that the oscillations react in a more unstable way than in
fundamental mode pulsators to small changes driven by stellar
evolution, as mentioned earlier. It would therefore not be surprising
if they reacted in a similar manner to the periodic approach of the
companion.

%%%%%%%%%%%%%%%%%%%%%%%%%%%%%%%%%%%%%%%%%%%%%%%%%%%%%%%%%%%%%%%%%%%%%
\section{Conclusions}
\label{sec:conclusions}
%%%%%%%%%%%%%%%%%%%%%%%%%%%%%%%%%%%%%%%%%%%%%%%%%%%%%%%%%%%%%%%%%%%%%

In this work, we have taken advantage of the large body of RV
measurements in the literature of the past 125~yr to revisit the
pulsation properties of Polaris, and to improve its spectroscopic
orbit. All of the spectroscopic elements now have uncertainties that
are several times smaller than before, and should serve as the basis
for a more accurate mass determination for the Cepheid, once
additional astrometric observations along the 30~yr orbit are secured.

Notwithstanding its long observational history, Polaris continues to
be an object of great astrophysical interest.  Our RV analysis has
provided clear evidence confirming earlier indications that the
pulsation period is now becoming shorter, starting around the year
2010. This change in direction, compared to the prevailing trend
during most of the 20th century, is inconsistent with the previous
interpretation that the period change reflected rapid evolution across
the instability strip.

Our analysis has also enabled us to more precisely map changes in the
RV amplitude of the pulsation since 1900. We have shown that after the
sharp $\sim$1960--1990 decline, and the almost equally rapid recovery
thereafter, the amplitude stopped increasing roughly around 2016. It
may have started to decrease again, although this needs to be
confirmed with further measurements in the coming years.

These irregular period and amplitude changes seem consistent with the
claim of \cite{Evans:2002, Evans:2018} that the oscillations in
overtone pulsators such as Polaris are more unsteady than those in
fundamental mode pulsators, and sensitive to small evolutionary
changes. Interestingly, we have found that these changes appear to
have a connection to the binarity of Polaris, always occurring near
the times of periastron passage in the 30~yr orbit, at least so
far. This could be the result of forced tidal perturbations by the
companion, and would support the idea by Evans and collaborators that
the pulsations are easily influenced by other effects, in this case an
external force.

\begin{acknowledgements}

We are grateful to D.\ Turner, the referee, for sharing the
spreadsheets with his original data compilations and reductions, and
for helpful comments on the original manuscript. We also thank
B.-C.\ Lee and R.\ B\"ucke for providing their unpublished radial
velocities for our analysis, and R.\ Anderson for sending his
photometric amplitude measurements based on the SMEI observations.

This research has made use of the SIMBAD and VizieR databases,
operated at the CDS, Strasbourg, France, and of NASA's Astrophysics
Data System Abstract Service.  The computational resources used for
this research include the Smithsonian High Performance Cluster
(SI/HPC), Smithsonian Institution (https://doi.org/10.25572/SIHPC).

\end{acknowledgements}

\section{Data Availability}

The data underlying this article are available in the article and in
its online supplementary material.

%%%%%%%%%%%%%%%%%%%%%%%%%%%%%%%%%%%%%%%%%%%%%%%%%%%%%%%%%%%%%%%%%%%%
\appendix
\label{sec:appendix}
\section{Historical Radial Velocities}
%%%%%%%%%%%%%%%%%%%%%%%%%%%%%%%%%%%%%%%%%%%%%%%%%%%%%%%%%%%%%%%%%%%%%

Here we provide details of the various sources of radial velocity
measurements for Polaris listed in Table~\ref{tab:historical.rvs} of
the main text, in the order in which they appear there. We describe
only the original sources, except where the velocities have been
adjusted or otherwise superseded in subsequent publications.

\vskip 5pt\noindent{\bf \cite{Vogel:1895}:} These are the earliest two
radial velocity measurements reported for Polaris, obtained at the
Potsdam Observatory (Germany) in November and December of 1888. They
were originally published in units of German geographical miles
(7.42~\kms), and are too poor to be of use here.

\vskip 5pt\noindent{\bf \cite{Roemer:1965}:} The longest and most
complete series of measurements, consisting of 1180 velocities made
with the Mills spectrograph on the 36-inch refractor at the Lick
Observatory (California, USA) between 1898 and 1958. These data cover
more than two cycles of the binary, including three periastron
passages, and are valuable for improving the orbital period when
combined with modern observations. We have assigned them initial
uncertainties of 0.5~\kms\ each. In the course of using these
measurements, we noticed a systematic shift in the residuals from a
preliminary orbital solution between the velocities before and after
about 1920. The ones before appear consistently more negative by about
0.8~\kms. While the reasons for this are unclear, the original
publication indicates that, prior to that date, no account was taken
of the flexure of the telescope tube, whereas the data taken afterwards
did include corrections for that effect. For the present analysis, we
have chosen to adjust the older velocities by +0.8~\kms, to place them
approximately on the same footing as the more recent Lick data. The
original publication also mentioned a hint of a semiregular variation
in the observations up to 1920, with a period of 6--8 yr and a total
amplitude of about 1~\kms. However, as this is not seen clearly in
subsequent data from the same series, we have not considered it
further. The study reported the first detailed determination of the
spectroscopic orbital elements, which had previously only been
published by others without associated uncertainties.

\vskip 5pt\noindent{\bf \cite{Frost:1899}:} Three observations from
the Yerkes Observatory (Wisconsin, USA) in August and September of
1899. They show clear variability due to the pulsations, but are too
few and too scattered to be useful. A fourth observation was mentioned
to have been taken, but the results were never published, as far as we
can tell.

\vskip 5pt\noindent{\bf \cite{Belopolsky:1900}:} These 17 velocity
measurements from the Potsdam Observatory were made between November
of 1899 and March of 1900, occasionally more than once per night. They
were gathered at a time when there was little change in the orbital
velocity of Polaris, which was at its maximum. They are therefore of
little use here for constraining the orbit, because of the need to
allow for an offset to account for the unknown velocity zeropoint
relative to other measurements.  Nevertheless, we have used them to
obtain a measurement of the pulsation amplitude and a time of minimum
velocity, to constrain changes in the orbital period.  Measurement
uncertainties were assumed here to be 1.5~\kms, based on the
intra-night scatter.

\vskip 5pt\noindent{\bf \cite{Hartmann:1901}:} This series of 35
measurements between March of 1900 and January of 1901 was also
gathered at the Potsdam Observatory. Individual uncertainties as
reported in the publication range from 1.8~\kms\ for the earlier
measurements to 0.7~\kms\ for the more recent ones.

\vskip 5pt\noindent{\bf \cite{Kustner:1908}:} Seven velocity
observations from the Bonn Observatory (Germany) were made in July of
1903 and May--June of 1905, at a time when the orbital velocity was
changing significantly.  Typical uncertainties as reported in the
paper are 0.95~\kms.

\vskip 5pt\noindent{\bf \cite{Abt:1970}:} The catalogue of Mount Wilson
Observatory (California, USA) radial velocity measurements lists only
3 observations made with different instrumental dispersions, two on
the same night in 1916, and the other in 1932. They are not useful for
the present work.

\vskip 5pt\noindent{\bf \cite{Henroteau:1924}:} This source contains
58 measurements of the RV made at the Dominion Observatory (Ottawa,
Canada) between April and September of 1923. The orbital velocity of
Polaris was not changing significantly during this time, so the
observations do not constrain the orbit for the same reason mentioned
earlier for the \cite{Belopolsky:1900} velocities. However, they do
provide useful measurements of the pulsation amplitude and time of
minimum velocity.

\vskip 5pt\noindent{\bf \cite{Schmidt:1974}:} These authors reported
only a single RV measurement of Polaris, obtained in 1971 at the Kitt
Peak Observatory (Arizona, USA). It is one of the very few
measurements we have located in the poorly observed 20~yr period
between 1960 and 1980, but it does not provide any useful information
for this work.

\vskip 5pt\noindent{\bf \cite{Wilson:1989}:} These three RVs of
Polaris from 1977, obtained at the McDonald Observatory (Texas, USA),
are the only others found between 1960--1980, but are also too few to
be useful. They cover less than one pulsation cycle.

\vskip 5pt\noindent{\bf \cite{ArellanoFerro:1983a}:} This series of 35
RV measurements was made at the David Dunlap Observatory and the
Dominion Astrophysical Observatory (DAO), both in Canada, from July of
1980 to January of 1982.  The uncertainties were adopted as published,
after a conversion from probable errors to mean errors. All but 6 of
these measurements were later republished by \cite{Kamper:1996}, along
with many others also from the David Dunlap Observatory. In that
publication, a +0.5~\kms\ offset was applied to the
\cite{ArellanoFerro:1983a} RVs to bring them onto the same scale as
the rest of those measurements. For the present work we have applied
the same +0.5~\kms\ offset to the \cite{ArellanoFerro:1983a}
velocities, including the 6 not listed by Kamper, so that they may be
considered jointly with the Kamper velocities and avoid adding another
free parameter to our analysis that would otherwise be needed to
account for the difference in the zeropoints.

\vskip 5pt\noindent{\bf \cite{Beavers:1986}:} This work reports a
single RV measurement made in 1980 at the Erwin W.\ Fick Observatory
(Iowa, USA).  We have ignored for the present work.

\vskip 5pt\noindent{\bf \cite{Kamper:1984}:} This is a small subset of
the velocities published later by \cite{Kamper:1996}, in which minor
adjustments were made to the measurements. We therefore consider them
to have been superseded by the latter publication (see below).

\vskip 5pt\noindent{\bf \cite{Kamper:1996}:} This extensive series of
421 observations at the David Dunlap Observatory (1983--1995) relied
on a variety of observational and measurement techniques, beginning
with photographic plates at several dispersions until early 1990, then
using a Reticon detector, and finally replacing it with a CCD and
analysing the spectra with cross-correlation methods.  These changes
in instrumentation and processing made it difficult to place all of
the velocities onto a consistent system, despite the use of IAU radial
velocity standards in some intervals, or of the telluric oxygen band
at 6300~\AA\ as the velocity reference, for the more recent CCD
data. As a result, the author cautioned that the earlier measurements
that did not use the oxygen band may be subject to systematic errors
of comparable size to the frame-to-frame scatter. Clear evidence of
this was shown in Figure~5 by \cite{Anderson:2019}, in which the raw
velocities from the Reticon observations are seen to be systematically
higher than other measurements taken near in time, by roughly 1~\kms.
A similar offset is seen for observations during the 1992 July--August
run, in the same direction, and for an entire campaign during 1991
July-September in which a fibre link was used. In this latter case the
shift appears to be about 1~\kms\ in the opposite direction (i.e.,
lower than other contemporaneous velocities). With some hesitation, we
have chosen here to adjust the measurements in those intervals by the
amounts indicated, in an attempt to remove these rather obvious shifts
and make the entire dataset more homogeneous. A few other intervals
with smaller systematic deviations are discernible, but we have
refrained from further tinkering with the data and simply accept those
instances as contributing to the natural scatter.  As mentioned
earlier, \cite{Kamper:1996} list all but 6 of the
\cite{ArellanoFerro:1983a} velocities, which we will not consider as
part of this group. Additionally, it lists 79 observations after 1994
April that have been superseded by a later publication
\citep{Kamper:1998}, in which thermal effects were corrected. For the
present work we will consider this dataset to contain only the $421 -
35 - 79 = 307$ measurements that are not reported by either
\cite{ArellanoFerro:1983a} or \cite{Kamper:1998}. This collection of
observations is important because it has complete coverage of the
periastron passage of 1987, from velocity minimum to maximum, and part
of the way down again. Uncertainties for the individual observations
depend on the technique, and from our best understanding of the
description, they range between about 0.5~\kms\ for the photographic
measurements and 0.15~\kms\ for the more recent observations recorded
in digital form.  These observations were used by Kamper, both
separately and in combination with the Lick observations of
\cite{Roemer:1965}, to derive the elements of the spectroscopic binary
orbit that have been most commonly used in subsequent investigations
of Polaris.

\vskip 5pt\noindent{\bf \cite{Dinshaw:1989}:} These authors reported
175 velocities for Polaris, made between 1987 and 1988 at the
University of British Columbia's Wreck Beach Observatory (Canada). The
uncertainties have been adopted here as described in that work, with
those of measurements that are the mean from 2--3 spectra being
reduced by $\sqrt{2}$. A slight complication with these observations
is that, rather than being absolute, they were measured relative to
the first observation. More importantly, the RVs show a long-term
trend that the authors attributed mostly to orbital motion of the
binary, although from their description there may also be a component
due to mechanical flexure of the telescope tube. This long-term trend
is shown in their Figure~2, together with a second-order polynomial
approximation that they then subtracted from the relative
velocities. The polynomial coefficients were not reported, and the
individual velocity measurements listed in their paper are only those
\emph{after} removal of that long-term trend.  For our analysis, we
digitised the figure to extract the coefficients of the quadratic fit,
and then removed it from the listed measurements in order to recover
the relative velocities. In addition to the RV changes due to the
4-day pulsation, \cite{Dinshaw:1989} reported the detection of a
signal with a 45 day period and a total amplitude of about
1~\kms. Prior to using these data for our orbital analysis, we
subtracted this variation out, as described in the main text.

\vskip 5pt\noindent{\bf \cite{Sasselov:1990}:} These authors reported
two RV measurements made at infrared wavelengths (1.1~$\mu$m) with the
Fourier transform spectrometer at the Canada-France-Hawaii Telescope
(Hawaii, USA). They were obtained 4 nights apart in September/October
of 1988, but are not useful for our purposes.

\vskip 5pt\noindent{\bf \cite{Gorynya:1992}:} This work reports 32 RVs
made in 1990 and 1991 at the Shternberg State Astronomical Institute
in Moscow (Russia), with a CORAVEL-type spectrometer delivering
typical precisions of 0.3~\kms.

\vskip 5pt\noindent{\bf \cite{Garnavich:1993}:} This short note
reports only that Polaris was monitored beginning in 1992 with a
hydrogen fluoride gas absorption cell on the DAO coud\'e spectrograph,
achieving internal velocity precisions of tens of m~s$^{-1}$ (higher
than any of the previous velocities). As far as we are aware, the
individual measurements, which appear to have been made over at least
a year, were never published or discussed in the literature, and are
not available.

\vskip 5pt\noindent{\bf \cite{Hatzes:2000}:} This series of 42 high
precision velocity measurements was made in 1992--1993 on the 2.1m
telescope at the McDonald Observatory (Texas, USA), using an iodine
gas absorption cell as the reference. These are again relative
velocities, with an internal precision estimated by the authors to be
50--70~m~s$^{-1}$. We adopted the higher value for our work. After
removal of the pulsation variation, a frequency analysis by
\cite{Hatzes:2000} revealed a 40-day periodicity in the residuals,
similar to the signal found by \cite{Dinshaw:1989}, but with a smaller
total amplitude of about 0.4~\kms. As in that case, we have removed
this variability before using these relative velocities to constrain
the binary orbit.

\vskip 5pt\noindent{\bf \cite{Gorynya:1998}:} These authors reported
an additional 40 velocities of Polaris made in 1994, with the same
instrumentation as their 1992 paper (see above). We consider them
together with those measurements for the purpose of updating the
spectroscopic orbit.  We adopted the velocity uncertainties as
published.

\vskip 5pt\noindent{\bf \cite{Usenko:2015}:} A total of 56 velocities
were made with three different telescope/instrument combinations in
the US and in Russia, between 1994 and 2009. The 2003--2004 subset of
these measurements shows little coherence, with the 4-day pulsation
variability not clearly present. The author pointed out the unusual
behaviour of the H$\alpha$ line during these years (asymmetries, and a
significant RV difference between the core of the H$\alpha$ line and
the RVs from metal lines). While this could indicate a real phenomenon
in the atmosphere of Polaris, much higher quality measurements during
2004 by \cite{Eaton:2020}, which we describe below, appear completely
normal and show the pulsations clearly. We are inclined to believe
that Usenko's 2003--2004 measurements may have been impacted by
significant fluctuations in the velocity zero point, and we have
elected not to use them here. The remainder of their observations
appear to be unaffected.

\vskip 5pt\noindent{\bf \cite{Kamper:1998}:} This dataset consists of
212 RV measurements of Polaris from the David Dunlap Observatory, made
between 1995 and 1997. Uncertainties were adopted as published. Of
these measurements, 79 had been reported earlier by
\cite{Kamper:1996}, but were corrected here for thermal effects. All
of these velocities were measured relative to the first one, and the
listing of the individual values in Table~1 of \cite{Kamper:1998} has
the orbital motion subtracted out, using the elements published by
\cite{Kamper:1996}. For our purposes, we have added the orbital motion
back in, using those same elements.

\vskip 5pt\noindent{\bf \cite{Eaton:2020}:} This is the second most
extensive set of RVs for Polaris, after the Lick series. The 679
observations, of excellent quality (typical internal errors of
0.11~\kms), were obtained between 2003 and 2009 with the Tennessee
State University's 2m Automatic Spectroscopic Telescope at the
Fairborn Observatory (Arizona, USA). A subset of these measurements
was used by \cite{Bruntt:2008} in a study of changes in the pulsation
period and amplitude. The remaining observations appear not to have
been used until now.

\vskip 5pt\noindent{\bf \cite{Lee:2008}:} This is another series of
very precise relative velocity measurements made with an iodine
absorption cell, using a high-resolution echelle spectrograph on the
1.8m telescope at the Bohyunsan Optical Astronomy Observatory
(Korea). Although the RVs were not listed in the paper, the lead
author was kind enough to give us access to the 265 measurements,
which have typical formal uncertainties of 10--20~m~s$^{-1}$. As shown
in Figure~1 by those authors, the velocities display a clear trend due
to the orbital motion of the binary. In addition to the well-known
pulsation, they reported an additional $\sim$120 day periodicity with
a peak-to-peak amplitude just under 0.3~\kms. We have removed this
signal in preparation for use of these velocities for our orbital
analysis, as described in the main text.

\vskip 5pt\noindent{\bf \cite{Bucke:2021}:} As with the previous
dataset, these RVs for Polaris have not been published, but were made
available to us by the author, Roland B\"ucke, who is an amateur
astronomer based in Germany.\footnote{See
  \url{http://astro.buecke.de/index.html}} This is an untapped series
of 296 observations made between 2005 and 2023, using a backyard
8-inch and later an 18-inch Dobsonian telescope, coupled with an $R
\approx 3500$ spectrograph via an optical fibre. The measurements are
relative to the first observation, and span more than 60\% of the
30~yr orbital cycle. They provide dense coverage of the phases between
minimum and maximum radial velocity, including the latest periastron
passage of 2016. The spectrograph is not thermally controlled, and the
velocities display obvious seasonal changes that we have removed with
a spline approximation (after subtracting the pulsation and orbital
variations).  Further details are given in the main text. Initial
uncertainties for our analysis were adopted as supplied by the author.

\vskip 5pt\noindent{\bf \cite{Fagas:2009}:} This is another dataset
that appears to have never been published. The scant information
available indicates that 330 spectra were obtained over a 7 month
period (2007 December to 2008 July), with the 0.4m Poznan
Spectroscopic Telescope (Poland) and a fibre-fed spectrograph. While
the original data are not available to us, the authors reported a
useful measurement of the total velocity amplitude of $2A_{\rm puls} =
2.52 \pm 0.03~\kms$, from the first four months of their
observations. This value is consistent with others obtained during the
same period (see the main text).

\vskip 5pt\noindent{\bf \cite{deMedeiros:2014}:} The brief report by
these authors indicated that 13 RV measurements of Polaris were made
over a period of 6098 days using the CORAVEL instrument on the 1.93m
telescope at the Haute-Provence Observatory (France). They were never
published, but in any case, because of the sparseness of the
observations, the measures are unlikely to be of much help for our
analysis.

\vskip 5pt\noindent{\bf \cite{Anderson:2019}:} This is a high-quality
set of 161 observations, obtained between 2011 and 2018 with the
Hermes spectrograph on the 1.2m Mercator telescope on the Roque de los
Muchachos Observatory (La Palma, Canary Island, Spain).  The
velocities are on an absolute scale, and have formal uncertainties of
15~m~s$^{-1}$, which reflect only the short-term precision. A template
matching procedure designed to remove the pulsation variations was
carried out by Anderson, and applied to the Hermes observations and
those of \cite{Kamper:1996} to derive a new set of orbital elements
for the binary.

\vskip 5pt\noindent{\bf \cite{Usenko:2016, Usenko:2017, Usenko:2018,
    Usenko:2020}:} These studies continued the RV monitoring of
Polaris initiated by \cite{Usenko:2015}, but now mostly with the 0.81m
telescope at the Three College Observatory (North Carolina, USA), and
occasionally with a 0.6m telescope at the Kernesville Observatory,
also in North Carolina. Collectively, the observations span the
interval 2015--2020 and cover the recent periastron passage of 2016. A
total of 21, 49, 67, and 53 velocity measurements were reported in
each of these four studies, respectively. In the final paper of this
series, the authors derived an updated spectroscopic orbit combining
their own observations with seasonal averages from
\cite{Anderson:2019} and \cite{Roemer:1965}.

\end{document}